\documentclass[preprint,12pt]{elsarticle}

\usepackage{amssymb}

\usepackage[nodots]{numcompress}

\usepackage[utf8x]{inputenc}
\usepackage{epsf}

\providecommand{\e}[1]{\ensuremath{\times 10^{#1}}}

\journal{Acta Materialia}

\begin{document}

\begin{frontmatter}



\title{Molecular Dynamics Simulations of Compression-Tension Asymmetry in Plasticity of Fe Nanopillars}


\author{Con Healy, Graeme Ackland}

\address{School of Physics, The University of Edinburgh, Edinburgh, EH9 3JZ, United Kingdom}

\begin{abstract}

Tension-compression asymmetry is a notable feature of plasticity in
bcc single crystals.  Recent experiments reveal striking differences
in the plasticity of bcc nanopillars for tension and compression.
Here we present results from molecular dynamics simulations of
nanopillars of bcc Fe in tension and compression.  We find that a 
totally different deformation mechanism applies in each cases:
dislocation glide in compression and
twinning in tension. This difference explains experimentally-observed 
asymmetry in the nanopillar morphology.

\end{abstract}

\begin{keyword}

Nanopillar \sep Plasticity \sep Molecular Dynamics \sep Dislocations \sep Twinning



\end{keyword}

\end{frontmatter}


\section{Introduction}


The development of micro-mechanical testing on single crystals
\cite{Uchic04} has enabled the study of plastic deformation to be
taken to the microscopic level.  Under uniaxial compression of single
crystals, yield can be observed on a single slip plane.  Samples can
now be tested in both tension and compression, and there is a marked
contrast between the shear banding in compression and twinning
observed in tension\cite{JangNatNano}. Recent work has shifted to bcc
materials, where again asymmetry is observed.  

For dislocation motion, FCC materials generally obey Schmidt's law: a
dislocation moves once the shear stress on its glide plane exceeds a
critical magnitude.  Thus Schmidt's law predicts identical behaviour
under tension and compression.  By contrast, BCC materials typically
do not obey Schmidt's law: dislocation behaviour depending on stresses
outside the slip plane.  This can be traced to the core structure of
dislocations.

Recent experiments by Kim and Greer revealed a tension-compression
asymmetry in the plasticity of bcc nanopillars \cite{KimGreer,
  KimJangGreer1, KimJangGreer2, KimJangGreer3}.  In compression,
they observe plastic deformation confined mostly to narrow slip bands, and
stress vs strain behaviour characterised by many strain bursts.
This stress vs. strain signature would suggest deformation by
dislocation nucleation and motion.  
In tension they observe necking,  and
the stress vs strain behaviour revealed larger periods of almost
constant flow stress.  Kim and Greer suggested that the
twinning-antitwinning asymmetry could be a cause for the
compression-tension asymmetry\cite{KimGreer}.  However, their
experiments cannot determine what processes were occurring at the
atomic level.  Molecular dynamics simulations can be used to study
such atomic level processes.

The dependence of the critical resolved shear stress (CRSS)
on the sense of shear (i.e. tension or
compression)  is a feature of bcc single
crystals \cite{TaylorElam1926, Christian1983, HollangSeeger}.  
Initial explanations for this phenomenon focused on 
twinning-antitwinning slip asymmetry.
$\{112\}\langle 111\rangle$ type twins in bcc crystals can be viewed as
$\frac{1}{2}\langle 111\rangle $ dislocations split into $\frac{1}{6}\langle 111\rangle $
fractional dislocations which spread out on subsequent planes to
create a twin plane \cite{hull}.  The movement of these fractional
dislocations is only permitted in one direction on a given $\langle 111\rangle $
axis in order to create a twin boundary.  Glide in the opposite
direction (the anti-twinning direction) creates an unstable stacking
fault which is not a twin plane.  The CRSS for glide of
$\frac{1}{6}\langle 111\rangle $ fractional dislocations in the twinning direction
is therefore lower. 

Work by Vitek \textit{et al} gives further weight
to this argument\cite{DuesberyVitekOverview,ItoVitek}:  In this study, the gamma surface for a $\{112\}$
plane was calculated using a Finnis-Sinclair potential, and an
asymmetry in the $\langle 111\rangle $ direction was found.  
For a crystal in a fixed orientation,
relative motion of subsequent slip planes in compression will always
be in the opposite direction to relative motion of subsequent slip
planes in tension. The directional asymmetry in the glide of
dislocations is consequently thought to result in a
compression-tension asymmetry for single crystals
\cite{DuesberyVitekOverview, Christian1983, KimGreer, KimJangGreer1,
  KimJangGreer2}.
Further study\cite{DuesberyVitekOverview,   GrogerVitek2, GrogerVitek3} revealed another mechanism for CRSS
dependence on the sense of shear.  Molecular dynamics calculations
showed that the core structure of dislocations in bcc materials was
altered by non-glide shear stresses.  Due to a three-fold symmetry in
the core structure of $\frac{1}{2}\langle 111\rangle $ screw dislocations in bcc,
the change in the core structure depends on the direction of the shear
stress.  Unlike the twinning-antitwinning asymmetry,
these changes in the core structure affect slip on all slip planes,
not just glide on $\{112\}$ planes. 

Previous molecular dynamics studies of plasticity in nanopillars of bcc Fe have focused on tensile strains.
Observed plasticity mechanisms for tension have varied, with some studies reporting deformation by dislocations and phase transitions while others report deformation by twinning.
These differences in plasticity behaviour may be due to a number of factors, including the potentials used and differing pillar geometries.
In particular, the value for the difference in energy per atom between the fcc and bcc phase of the material can differ significantly across different potentials.
Using a potential where this energy value is low may result in a tendency for bcc Fe to change to fcc in certain parts of the system under the influence of an applied stress or strain.
The crystallographic orientation of the surfaces of the pillars may also affect the simulation results.
It is energetically favourable for bcc structures such as nanopillars to have low energy $\langle 110 \rangle$ surfaces.
Pillars containing high energy surfaces may have a tendency to deform in a way which allows surface reconstructions to occur as reported by Ackland in fcc nanopillars \cite{MOLDYPaper}.
The direction of the compression or tension could also affect the plasticity behaviour as the tendency to observe dislocation glide in a material depends on the schmidt factors for 
the slip planes in the pillars and these schmidt factors in turn depend on the direction of the applied strain.
Here we offer a brief overview of previous molecular dynamics studies of plasticity in Fe nanopillars.

Zhang \textit{et al} \cite{Zhang2012164} performed simulations of tensile strain of nanopillars oriented along $\langle 100 \rangle$ and $\langle 110 \rangle$ directions.
The $\langle 100 \rangle$ pillars contained $\{ 100 \}$ type side surfaces and the $\langle 110 \rangle$ pillars contained two $\{ 100 \}$ and two $\{ 110 \}$ type side surfaces.
Deformation was observed to occur by dislocation glide in the $\langle 100 \rangle$ pillar and by a phase transformation from bcc to fcc in the $\{ 110 \}$ pillars.
This dependence of plasticity behaviour on pillar geometry may be due to the different crystallographic surface orientations in the side faces of these two pillars as well as 
differing schmidt factors for slip planes in each pillar.
The potential used in this study was one for Fe developed by Mishin \textit{et al} \cite{MishinMehl}.
For this potential, the difference in energy per atom between the fcc and bcc phases is ~50 meV  \cite{ MishinMehl}.
This value is lower than calculated in many \textit{ab initio} calculations which typically give values between 70 meV and 120 meV \cite{SandovalPotStudy, ironpot, PhysRevB.45.8887, PhysRevB.46.1870}.

Simulations of tensile strain of bcc Fe nanowires were also carried out by Sandoval and Urbassek \cite{Sandoval1}.
They constructed pillars oriented along $\langle 111 \rangle$ with a circular cross section.
In this study deformation was found to occur by a phase tranformation from fcc to a combination of fcc and hcp.
The potential used in this study was one for Fe developed by Meyer and Entel \cite{MeyerEntel}.
The energy difference in energy per atom between the fcc and bcc phases is ~40 meV for this potential \cite{SandovalPotStudy, MeyerEntel} which is again rather low in comparison to
values calculated in \textit{ab initio} calculations.

Li \textit{et al} \cite{Li_etal_nanowires} performed simulations of tensile strain on $\langle 100 \rangle$ pillars with $\{110 \}$ type side surfaces.
Deformation was found to occur by twinning.
The potential used in this study was one developed by  Mendelev \textit{et al}\cite{ironpot} for which the difference in energy per atom between the fcc and bcc phases is ~120 meV.

The phase transitions observed in the simulations by Zhang \textit{et al} and Sandoval and Urbassek are most likely the result of using potentials with low values for the 
difference in energy per atom between the fcc and bcc phases.
This is evidenced by the lack phase transitions in the simulations by Li \textit{et al}.
In this study, we use a modified version of the Mendelev \textit{et al} potential as the difference in energy per atom between the fcc and bcc phases lies within the range of values typically 
found from \textit{ab initio} calculations.
We construct nanopillars containing low energy $\{ 110 \}$ type side surfaces to avoid possible surface reconstructions which could
 occur in order to generate low energy surfaces on the pillars as the strain is applied.


One might speculate that the observed tension-compression asymmetry in
bcc nanopillars is related to the intrinsic non-Schmidt behaviour of
the dislocations.  Alternately, one might argue that the mechanism
should be the same as for fcc materials.  Here we use molecular
dynamics simulations to simulate the tension-compression asymmetry in
nanopillars of bcc Fe.  We show that the asymmetry is due to different
deformation mechanisms: dislocation glide in compression and twinning
in tension.  For completeness, we also examine fcc nanopillars and find
no such effect.

\section{Simulation Details}

Our system consisted of a pillar with a square cross-section in
between two "indenter plates" as seen in figure
\ref{fig:compressionImages}(a).  The pillars had approximate
dimensions of 5.8$\times$5.8$\times$15.4 nm and contained
45513 atoms. 
The sample is confined by the indenter plates, each containing 9901 atoms
(12.1 nm $\times$ 12.1 nm $\times$ 0.9 nm).
Movement of indenter plate atoms is constrained in order to apply
external forces.  The side faces of the pillar were $\{110\}$ type
faces and the pillars were compressed in the [001] direction.  It is
necessary to have $\{110\}$ faces on the pillar as these are the
lowest energy surfaces.  Creating a pillar with higher energy surfaces
will result in recrystallization during the simulation and this has
been shown to have create spurious deformation behaviour\cite{MOLDYPaper}.

Prior to loading, all pillars were heated to 300K by running a
simulation for 50 picoseconds with a Nos\'{e}-Hoover
thermostat\cite{nose, Hoover}.  Uniaxial strain was applied by moving
the indenter plates and rescaling the atomic coordinates in the
direction of loading by 0.05\% at 2 picosecond intervals.  The
resulting strain rate is 2.5\e{8} $s^{-1}$.  This method of
compressing the system, by rescaling the coordinates of the atoms in
the compression direction, is required so that a shock wave is not
produced in the pillar.  Since the thickness of the pillar may change
through the simulation, ``stress'' on the sample is not easily
defined: we measure the force required to hold the indenter plate in
position, converted to a stress by dividing by the indenter area.
Simulations were carried out using the MOLDY molecular dynamics
code\cite{MOLDYPaper}. The potential function used was that developed
by Hepburn and Ackland \cite{IronPotential}.

Images were created using AtomEye\cite{atomeye} some analysis of dislocations was carried
out using the OVITO package \cite{ovito_ref}.

\section{Results}

\subsection{Compression}

\begin{figure}[htb]     
        \begin{center}
          \epsfxsize=100mm         
          \epsffile{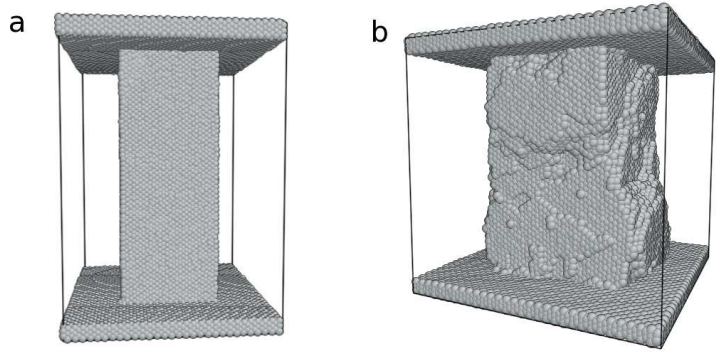}
\end{center}
\caption{(a) Pillar prior to applied strain, with the atoms in the plates at the top and bottom of the pillar constrained. (b) Pillar following a compressive strain of 26\%}
\label{fig:compressionImages}                 
\end{figure}

\begin{figure}[htb]     
        \begin{center}
          \epsfxsize=75mm         
          \epsffile{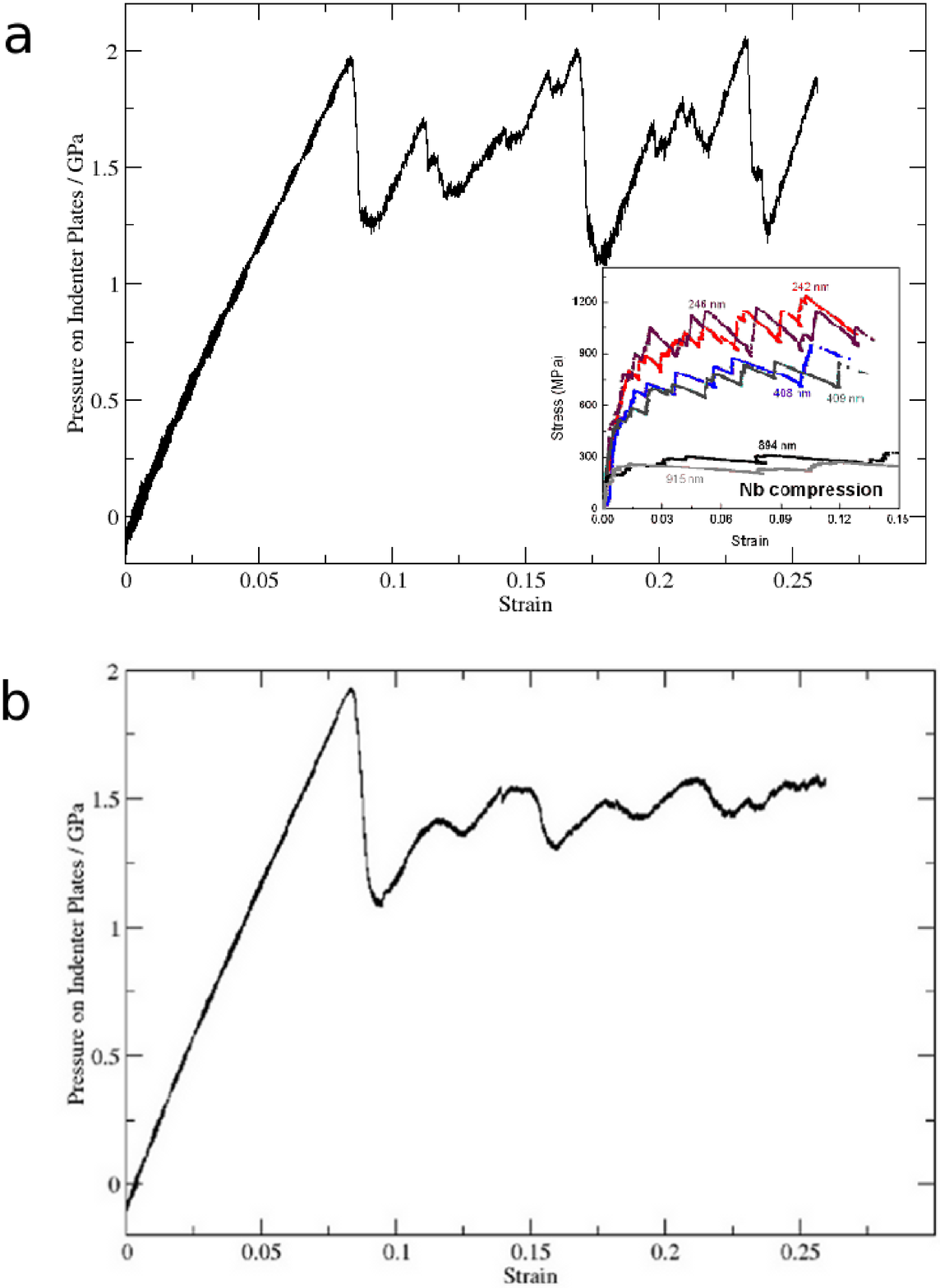}
\end{center}
\caption{(a) Plot of pressure on indenter plates vs strain for pillar
  in compression. The inset image shows a graph of Stress vs Strain
  for Nb micropillar compression experiments under stress control 
performed by Kim  \textit{et al}\cite{KimJangGreer2}. 
The pressure on the indenter
  plates does not start at zero due to surface tension. (b) Average
  pressure on indenter plates vs strain over 60 pillar compression
  simulations.} 
\label{fig:compressionPlots}                 
\end{figure}

Under compressive strain-loading, deformation in the pillars was mediated
mainly by dislocation glide.  Dislocations are created at the surfaces
of the pillars when the stress is high and move quickly through the
pillar.  
Typically, $\langle 111\rangle$ dislocations form at corners, which allows them to be short but 
requires a mixed edge-screw character.
This gives characteristic bursts of deformation
leading to sudden decreases in stress, as can be seen in the graph
in figure \ref{fig:compressionPlots}(a).  Each sudden drop in
pressure on the indenter plate happens when a dislocation is created
and moves through the pillar, creating a discrete strain burst. 
This behaviour is qualitatively similar to stress vs strain signatures obtained by Kim  \textit{et al}\cite{KimJangGreer2} in experimental studies of nanopillars of Nb,
 which can be seen in figure \ref{fig:compressionPlots} (b).
  Each dislocation slip event creates a step on the surface of the pillar.
 Figure \ref{fig:compressionImages}(b) shows an image of a
pillar following a strain of 26\%.  Many steps in the surface of the
pillar due to dislocation glide activity can be seen in this figure.


Initially, the flat side surfaces provide the only nucleation sites, and
yield depends on dislocation nucleation from perfect $\{110\}$ side surfaces.
At the initial yield corresponding to the first peak in figure \ref{fig:compressionPlots}(a), several dislocations are nucleated simultaneously.
Subsequent yield events generally involve the nucleation and glide of a 
single dislocation.

\begin{figure}[htb]     
        \begin{center}
          \epsfxsize=100mm         
          \epsffile{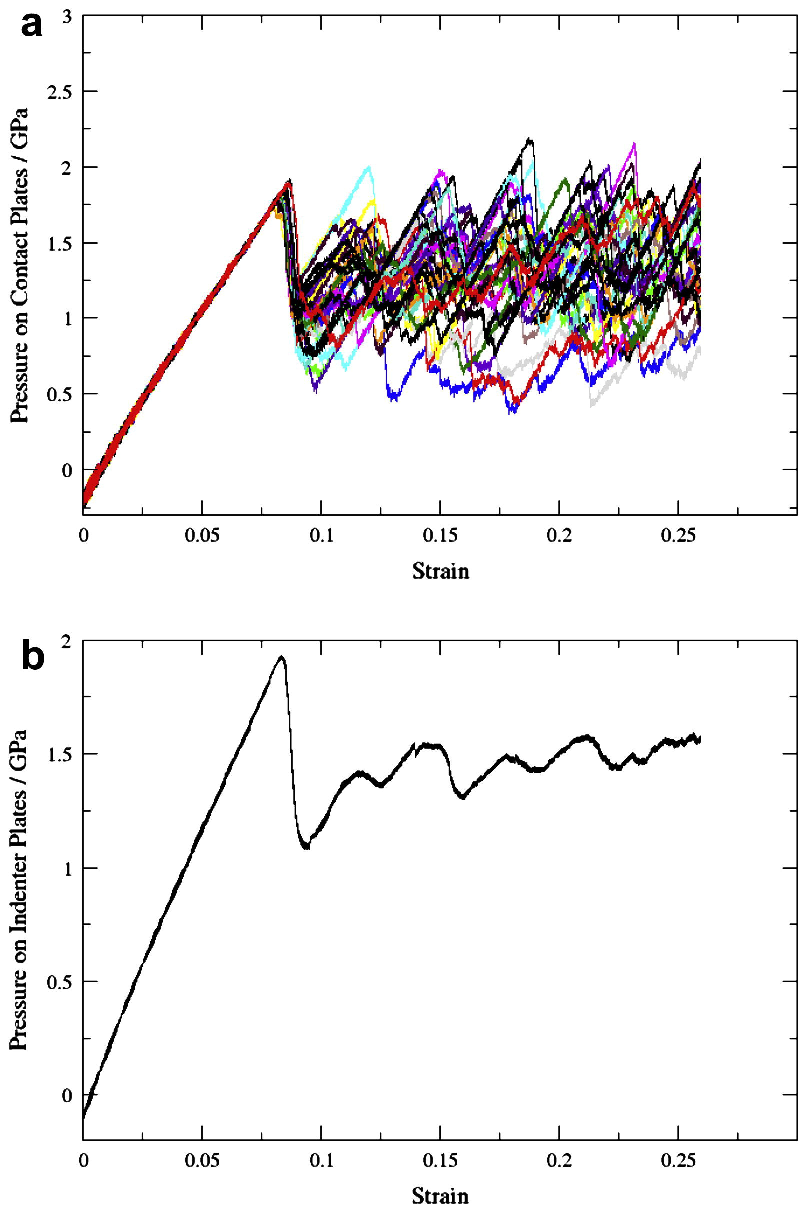}
\end{center}
\caption{(a) Indenter plate pressure vs strain for 60 different
  simulation runs superimposed on a single graph.  (b) Average
  pressure on indenter plates vs strain.  The error on the mean over
  60 simulations can be estimated at $\pm 0.13GPa$: apart from the
  first peak and recovery the ``features'' in this graph are consistent
  with statistical noise.
}
\label{fig:avgOf60}                 
\end{figure}

To whether this sequence of discrete yield events is deterministic or
random, we repeated the compression simulation 60 times.  In all
simulations the first dislocation creation event occurs at an indenter
pressure of 2GPa and at a strain of about 7.5\%.  Many dislocations
are created almost simultaneously at this point, and the drop in
stress varies by a factor of 2.  The stress then builds elastically to
a second yield event. Subsequent dislocation creation and glide events
occur stochastically within a range of indenter pressure values and at
varied values of strain.  This is illustrated by figure
\ref{fig:avgOf60}.  Figure \ref{fig:avgOf60} (a) shows plots of
indenter plate pressure vs strain for 60 different simulations
superimposed on one graph.  From this graph, it can be seen that the
indenter plate pressure and strain values at the initial yield point
are the same for all simulation runs.  These values are distributed at
random for subsequent dislocation glide events.  This interpretation
is confirmed by the average indenter plate pressure vs strain graph
shown in figure \ref{fig:avgOf60} (b) which shows the pressure on the
indenter plates at various strain values averaged over the 60
different pillar compression simulations.  Averaged
over many simulation runs, the indenter plate pressure approaches an
approximately constant value at large strain due to the random
positions of the stress peaks associated with each dislocation glide
event.

 Comparing with nanopillar experiments we see similar bursts of
 deformation: in our {\it strain} controlled simulations this
 manifests as a sharp drop in stress at fixed strain, whereas under
 {\it stress} control the same mechanism would give sharp increase in
 strain at fixed stress.  The experimental boundary conditions mix
 stress and strain control, hence the same mechanism manifests as
 simultaneous strain increase and stress drop.  The lower stresses
 required to initiate subsequent bursts is due to the roughening of
 the pillar having lowered the potential energy barrier for
 dislocation creation at the surface.  

\begin{figure}[htb]     
        \begin{center}
          \epsfxsize=100mm         
          \epsffile{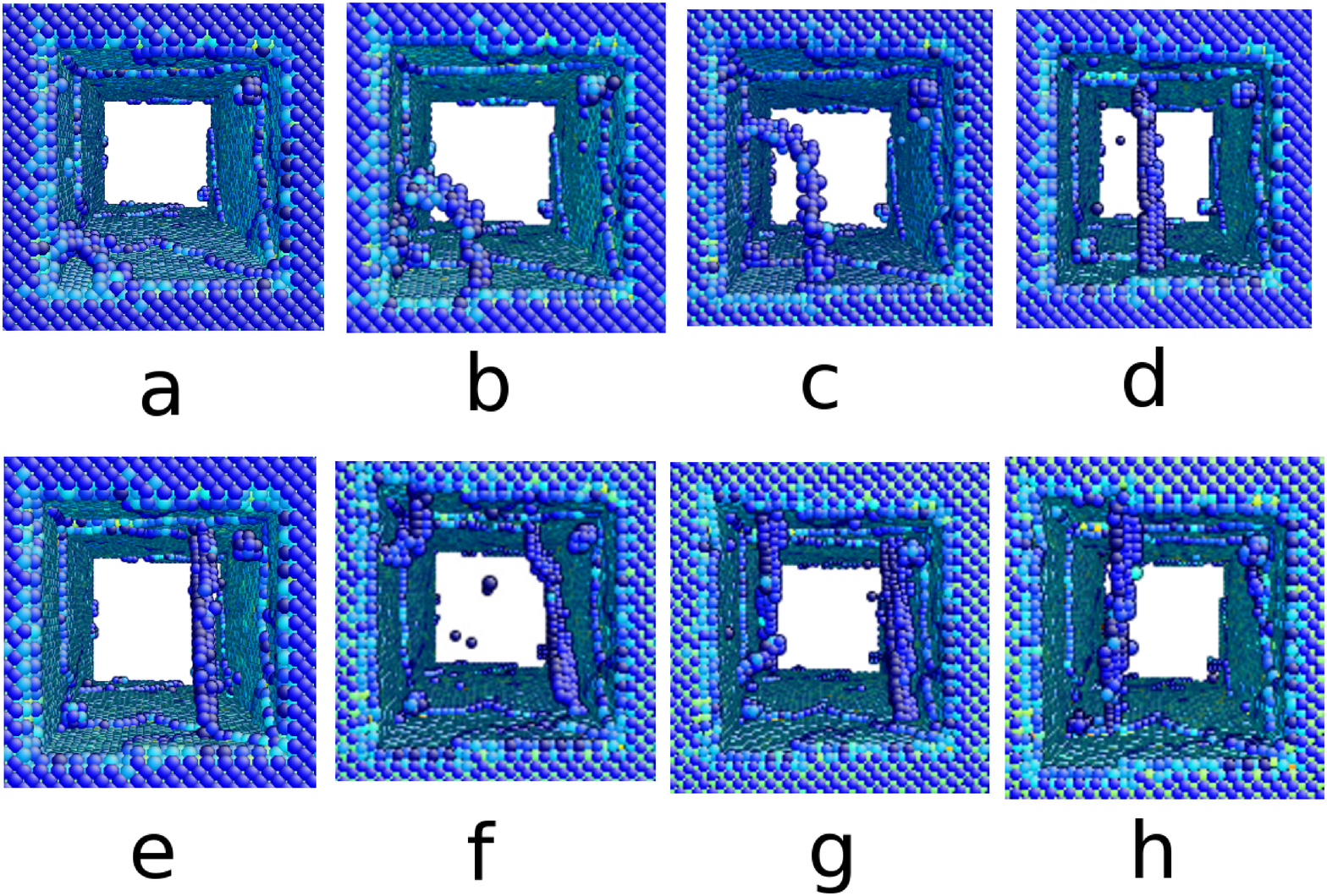}
\end{center}
\caption{View from the top of the pillar as a dislocation runs through the pillar in a compression simulation. Only atoms with high centrosymmetry parameter 
(typically free surface and dislocation line) are shown. Frames (a)-(d)
show a $\frac{1}{2}\langle 111\rangle $ dislocation emerge from the top left corner of the pillar. The dislocation initially has mixed character. However by frame (d) the dislocation forms a straight line
with pure screw character. Frames (e)-(h) show this dislocation move through the pillar until it meets one of the sides. A second dislocation can be seen emerging from another 
corner from frame (f) onwards.
 Although our simulations have a higher strain rate than the experiment,
this sequence shows that the deformation on compression is still primarily due to 
independent single-dislocation events.
}
\label{fig:compressionSequence}                 
\end{figure}

The sequence of images in figure \ref{fig:compressionSequence} shows a
typical occurrence of dislocation glide.  $\frac{1}{2}\langle 111\rangle $
dislocations are created at the corners of the pillars.  Initially the
dislocation lines are curved and have a mixed edge/screw character.
As the dislocation lines grow in length the dislocation line becomes
straight and the dislocation has pure screw character.  Each
dislocation glides on many different $\{110\}$ planes as a high degree
of cross slip occurs.  The dislocations continue to glide and
cross-slip along the direction of highest shear stress until they reach
one of the pillar surfaces, where they leave a step behind.  The
process then repeats.

\subsubsection{Dislocation Geometry}

The compression mechanism is clearly dislocation-based. Using the
\emph{dislocation extraction algorithm} (DXA) developed by Stukowski
and Albe \cite{DXA_ref} we find that all dislocations have
$\frac{1}{2}\langle 111\rangle $ Burgers vectors.
Dislocations could also be identified by measuring centrosymmetry of
nearest and second nearest neighbours of atoms with respect to the
atom in question.  A numerical measure of this central symmetry can be
found using the centrosymmetry parameter defined by Kelchner \emph{et
  al} \cite{centrosymmetry}.

\subsubsection{Atomic Shear Strain Analysis and Slip Planes}

\begin{figure}[htb]     
        \begin{center}
          \epsfxsize=30mm         
          \epsffile{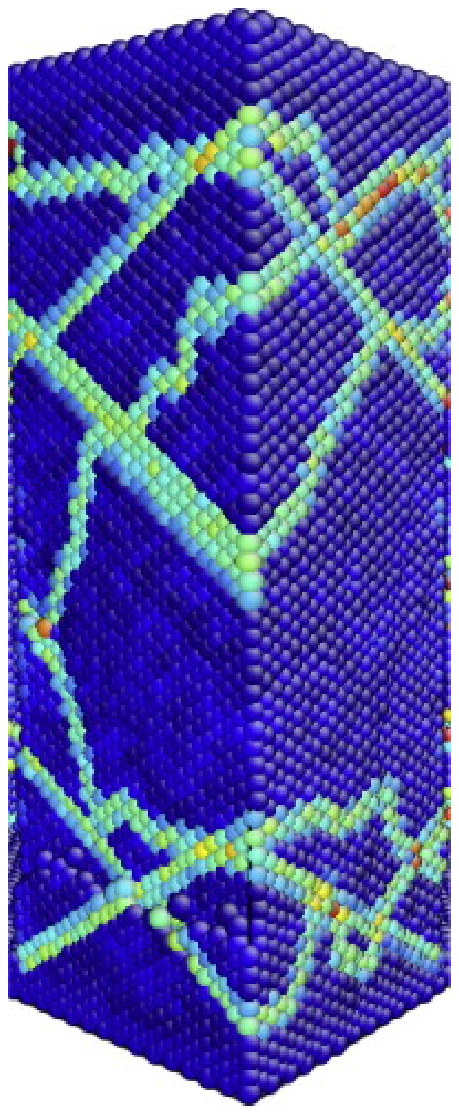}
\end{center}
\caption{Cross section of the pillar following a strain of 14\%. Atoms are coloured according to atomic shear strain. The light blue and green atoms have high atomic shear
strains while the dark blue atoms have low atomic shear strains.} 
\label{fig:atomicStrain}                 
\end{figure}

Atomic strain tensor analysis shows that a significant amount of cross slip occurs in the sample under compression.
Slip was found to occur on a series of connected $\{110\}$ planes for each dislocation.
Figure \ref{fig:atomicStrain} shows a cross section of the pillar following a compressive strain of 14\%.
Atoms are coloured according to the local atomic shear strain metric as defined by Shimizu \textit{et al} \cite{atomicstrain}.
Cross sectional slices were taken across two faces of the pillar and the viewing direction is a $\langle 100 \rangle$
 direction which runs diagonally through the centre of the pillar.
Atoms coloured light blue and green have high local atomic shear strains and are located on slip planes on which dislocation glide has occurred.
The jagged profile of these slip plane atoms indicates that a significant amount of cross slip has occurred.
Closer inspection of the slip plane atoms reveals that their profile is jagged on one side of the pillar and generally straight on the other side.
This is due to the orientation of the dislocations being perfect straight screw dislocations as seen in figure \ref{fig:compressionSequence}.
The straight lines formed by the dislocation lines were always parallel to one of the side faces of the pillar and therefore
 the slip path cross sections seen in figure \ref{fig:atomicStrain} must always be straight on one face of the pillar cross section even when a significant amount of cross slip
has occurred.
Cross slip allows many dislocations to traverse a path which runs at approximately $45\,^{\circ}$ to the direction of compression. 
The dislocations cross-slip so as to travel in this direction of maximum shear stress.

\subsection{Tension}

We have also carried out simulations on pillars in tension, and found that deformation occurs by twinning. 

This is consistent with
the mechanism whereby partial dislocations with burgers
vector $\frac{1}{6}\langle 112\rangle $ move on successive planes creating a twin
boundary on a $\{112\}$ plane\cite{hull} as
reported in similar molecular dynamics simulation recently by Li
\emph{et al} \cite{Li_etal_nanowires}. However,
in our simulations the twin formation is very rapid, and it
is not possible to identify individual fractional dislocation
events. 

\begin{figure}[htb]     
        \begin{center}
          \epsfxsize=50mm         
          \epsffile{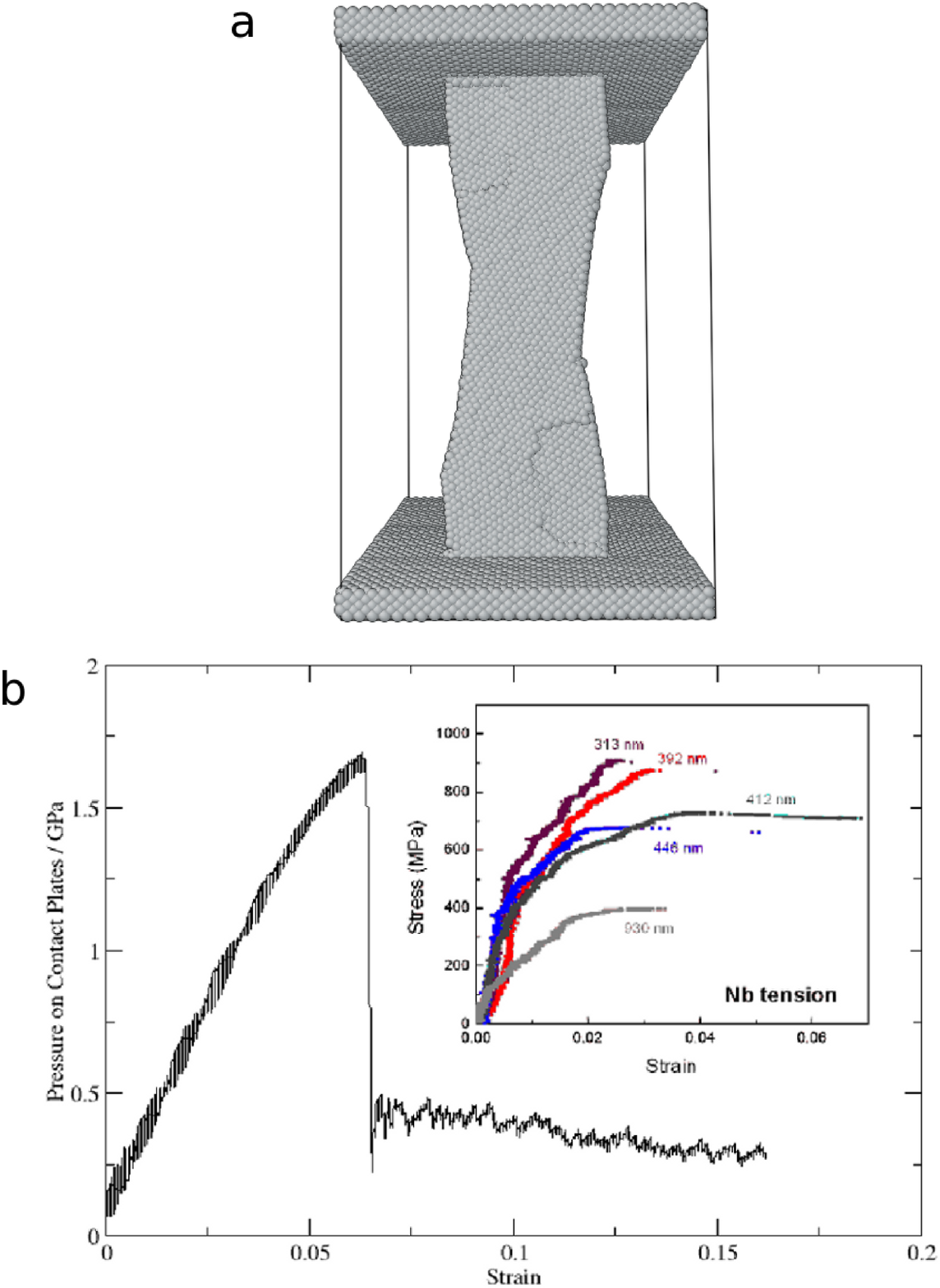}
\end{center}
\caption{ Image of pillar following a tensile strain of 16\%.}
\label{fig:tension16}                 
\end{figure}

A typical image of a pillar following a tensile strain of 16\% is
shown in fig \ref{fig:tension16}.  Twin formation occurred in a
single event - as shown in the massive yield event at 6\% strain in
\ref{fig:tension16}(b). 
After the initial yield event, twinning deformation proceeds by motion
of the twin boundary in a continuous process as shown by the plot of
indenter plate pressure vs strain in figure \ref{fig:tension16}(b).  A
large stress is required to create the twins initially, but the strain
subsequently continues at an almost continuous stress level as the
twin boundaries move.  The asymptotic stress is lower than that required
for compression by a factor of 3: more than can be accounted for by
changes in the cross-sectional area.
With the exception of the initial stress peak in \ref{fig:tension16}(b), this stress vs strain behaviour is qualitatively similar to the stress vs strain signature 
recorded in experiments on nanopillars of Nb reported by Kim \textit{et al}\cite{KimJangGreer2} which can be seen in figure \ref{fig:tension16}(b, inset).
The initial stress peak observed in our simulations is probably due to the high strain rates which we are limited to in molecular dynamics simulations.


\begin{figure}[htb]     
        \begin{center}
          \epsfxsize=100mm         
          \epsffile{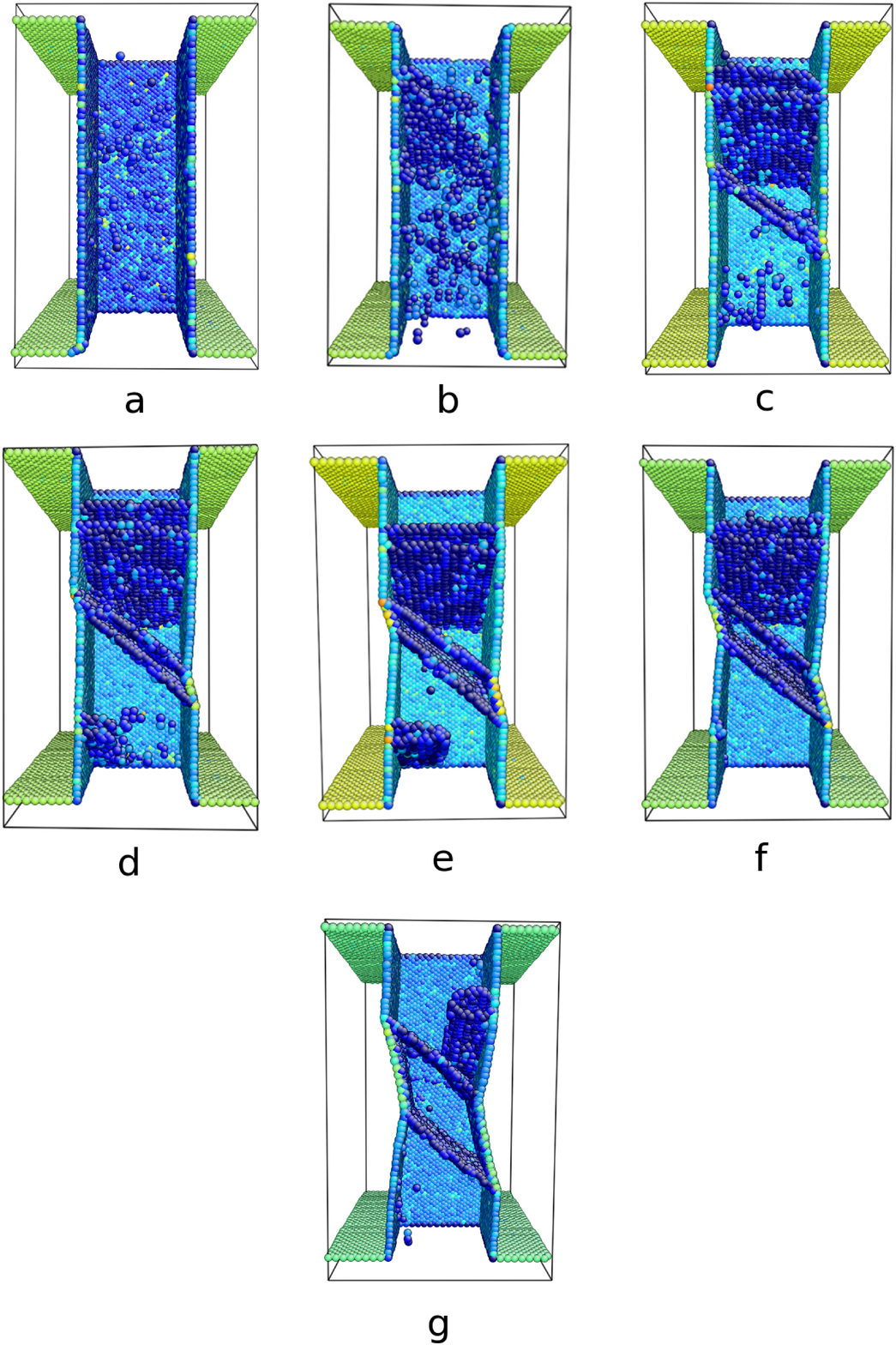}
\end{center}
\caption{Side view cross section of the pillar during tension. Only
  atoms with high centrosymmetry parameter are shown. Frames (a)-(f)
  show the creation of three twins in the initial yielding event
  following 7\% tensile strain. As deformation proceeds, two of these
  twins disappear until only the single, central, twin remains.
This can be
  seen in frame (g) which shows the pillar following 11\% tensile
  strain.}
\label{fig:tensionSequence}                 
\end{figure}

The sequence of images in figure \ref{fig:tensionSequence} shows the creation of pairs of twin boundaries during the tensile strain simulation.
When the pillar first begins to deform plastically, three pairs of twin boundaries are created.
As the separation between the twins in the middle of the pillar grows, the other two pairs of twins shrink in size to allow the central twins to move.

It should be noted that this twinning process results in the creation
of $\{100\}$ surfaces on the pillar in the plastically deformed region
in the centre.  These surfaces are unstable and our nanosample will
recrystallize if held at this strain and allowed to
relax. Alternately, if the stress is removed the sample will untwin in
order to reduce the surface energy.  In a sense, this nanopillar
exhibits superplasticity.  To study the deformation, we worked at
shear rates where these diffusive reconstructions did not occur.  This
is consistent with previous MD calculations.  Suppressing these
reconstruction events is justifiable because the driving force for
them is the dominant effect of surface energy, something which is only 
the case at the nanoscale, and not in pillars of the size considered 
in experiment.

\subsection{FCC materials}
We have tested the plasticity behaviour for fcc Cu nanopillars and
have found little differences between plasticity behaviour in
compression and tension.  Deformation in these pillars occurred
through partial dislocation glide on $\{111\}$ planes, without any cross slip
occurring, for both tension and compression.  The yield stress in
tension for these pillars was about half that in compression. This is
consistent with previous reports of MD simulations of plasticity in
copper \cite{Brown2010886, PhysRevB.79.075444}.

\section{Discussion}

Under compression, our Fe pillars deform by glide of
$\frac{1}{2}\langle 111\rangle$ dislocations on slip planes.  Under tension a
twinning mechanism operates.
This is consistent with much experimental evidence including that of
Kim and Greer \cite{KimGreer, KimJangGreer1, KimJangGreer2,
  KimJangGreer3} which showed bursts of dislocation motion punctuated
by periods of rising stress.

There are some differences in the behaviour between our MD simulations
and experimental results, due to differences in boundary conditions.
Experimental results for compression of bcc nanopillars often reveal
strain bursts in the stress vs strain behaviour.  In contrast we see a
series of sudden stress drops.  This apparent difference is because we
apply fixed strain, which means that events involving large
spontaneous strain are impossible.  There may be some difference due
to the small size of our pillars, the high strain rates which we have
to work with in MD: in particular the surface-driven recrystallization
may well be a size effect.  The experimental boundary conditions are
difficult to characterize: although nominally under stress control, they
actually show large sudden changes both in stress and strain.


In the tensile case, the twinning we observe is a possible mechanism
by which the necking seen in Kim and Greer’s experiments could occur.
The long periods of constant flow stress observed in these experiments
agrees well with that seen in simulations as seen in figure \ref{fig:tension16}.  The biggest
difference between simulations and experiment in this case is the
large initial stress required to nucleate the twin in our
simulations. This is because unlike the experiments we have an
atomically-flat surface, on a dislocation free sample.  Such high
yield stresses are observed in experiments with dislocation-free  
single-crystal whiskers.

A possible explanation for the observed asymmetry comes from the fact that dislocation cores in
bcc are spread in three dimensions.  The more compact the core, the
easier it is for the dislocation to move.  Hence, under compression, the core
will be compacted, favouring the dislocation mechanism, while under tension 
the core can  spread even further, making dislocation motion more difficult\cite{DuesberyVitekOverview,  GrogerVitek2, GrogerVitek3}.

The twinning-antitwinning asymmetry is not present in fcc crystals as
formation of twins is not mediated by fractional dislocations as in
bcc materials.  
The change in the core structure under stress in bcc
materials is dependent on the presence of edge components in
fractional dislocations\cite{DuesberyVitekOverview}.  Due to the lack
of fractional dislocations in fcc materials, this effect is not
observed in fcc materials. 

\section{Conclusion}

In bcc nanopillars, the deformation mechanism is completely different
in compression and tension.  This
asymmetry is not present in fcc nanopillars.

In the compressive case, plastic deformation is mediated by
dislocations.  Plastic deformation occurs in discrete bursts when one
or more dislocations are created at the pillar surfaces and move
through the pillar by glide.  A large amount of cross slip is observed
in this deformation regime allowing for the dislocations to move in
the direction of maximum shear stress.  By contrast, in the tensile
case deformation occurs through the creation and motion of twin boundaries.
This is consistent with asymmetry observed in various experiments.









\end{document}